\documentclass[twocolumn,showpacs,preprintnumbers,amsmath,amssymb,prb,superscriptaddress]{revtex4-1}
\usepackage{amsfonts}
\usepackage[english]{babel}
\usepackage[T1]{fontenc}
\usepackage{times}
\usepackage{mathrsfs}
\usepackage{graphicx}
\usepackage{dcolumn}
\usepackage{bm}
\usepackage[colorlinks,bookmarks=true,citecolor=blue,linkcolor=red,urlcolor=blue]{hyperref}
\usepackage[tight, FIGTOPCAP, hang, raggedright, nooneline]{subfigure}
\usepackage{hyperref}

\begin{document}

\title{Bulk-edge correspondence in fractional Chern insulators}
\author{Zhao Liu}
\affiliation{Beijing Computational Science Research Center, Beijing, 100084, China}
\author{D.~L.~Kovrizhin}
\affiliation{T.C.M. Group, Cavendish Laboratory, J. J. Thomson Avenue, Cambridge CB3 0HE, United Kingdom}
\affiliation{Imperial College London, London, SW7 2AZ, United Kingdom}
\affiliation{Russian Research Centre, Kurchatov Institute, 1 Kurchatov Sq., 123098, Moscow, Russia}
\author{Emil J.~Bergholtz}
\affiliation{Dahlem Center for Complex Quantum Systems and Institut f\"ur Theoretische Physik, Freie Universit\"at Berlin, Arnimallee 14, 14195 Berlin, Germany}

\date{\today}

\begin{abstract}
It has been recently realized that strong interactions in topological Bloch bands give rise to the appearance of novel states of matter. Here we study connections between these systems -- fractional Chern insulators and the fractional quantum Hall states -- via generalization of a gauge-fixed Wannier-Qi construction in the cylinder geometry. Our setup offers a number of important advantages compared to the earlier exact diagonalization studies on a torus. Most notably, it gives access to edge states and to a single-cut orbital entanglement spectrum, hence to the physics of bulk-edge correspondence. It is also readily implemented in the state-of-the-art density matrix renormalisation group method that allows for numerical simulations of significantly larger systems. We demonstrate our general approach on examples of flat-band models on ruby and kagome lattices at bosonic filling fractions $\nu=1/2$ and $\nu=1$, which show the signatures of (non)-Abelian phases, and establish the correspondence between the physics of edge states and the entanglement in the bulk. Notably, we find that the non-Abelian $\nu=1$ phase can be stabilized by purely on-site interactions in the presence of a confining potential.
\end{abstract}

\maketitle

{\it Introduction.~}Fractional quantum Hall (FQH) states \cite{laughlin83,mr} provide examples of some of the most unusual phases of matter supporting excitations with a fraction of the electron charge that obey anyonic statistics.\cite{fstat,fstatqh} One of the remarkable features of these states is that they realize a condensed matter example of the holographic principle,\cite{holography} which in this context defines the relation between the physics of gapped bulk and of gapless edge states at the sample's boundary.\cite{mr,wenedge}

Recently there has been a growing interest in lattice models harboring nearly dispersionless (flat) bands with nonzero Chern numbers, so-called fractional Chern insulators (FCIs),\cite{chernins1,chernins2,chernins3,cherninsnum1,cherninsnum2,c1c,c1d,c2,max,dipolar,kapit,moessner,beyondL,andreas,cooper,nonab1,yfwang,nonab2,nonab3,qi,grushin} which show states similar to those of the FQH effect. These novel systems do not require external magnetic fields and can potentially be realized at room temperature due to shorter lattice length scales compared to the typical magnetic length in quantum Hall (QH) systems. While the connections between many-body correlations, quantum entanglement, and the properties of edge states in the FQH case are relatively well understood, \cite{LiH,bulkedge1,bulkedge2} the corresponding physics in the FCIs has been less studied.

The standard numerical tool in the FCIs is exact diagonalization (ED) of small systems.\cite{cherninsnum1,cherninsnum2} Despite its considerable success in finding some of the robust FCI states there are a number of reasons to look for alternative approaches which would allow for a detailed understanding of larger systems. A potentially powerful method in this context is the density matrix renormalization group (DMRG);\cite{white} initially designed for strongly correlated one-dimensional systems, it has been successfully applied in the simulations of a variety of two-dimensional states of matter including geometrically frustrated magnets \cite{kagomesl,balents,depenbrock} and FQH systems.\cite{shibata,bk,feiguin,kovrizhin,jzhao,hu}

From modern developments in the theory of quantum entanglement, which brought new ideas of the area law,\cite{gammatorus} the entanglement spectrum,\cite{LiH} and matrix product states,\cite{zaletel,estienne,ttmps,dubail} it has now become clear that a cylinder geometry plays a very special role in the DMRG studies of strongly correlated systems.\cite{zaletel,frank} Recently two pioneering papers \cite{balents,cincio} suggested using this geometry to extract information about topological properties of correlated states, including the FCIs.

In this paper we consider a setup which generalizes the FCI description on a torus to the case of finite cylinders, by constructing the interaction matrix elements in the gauge-fixed version \cite{gaugefixing,wannier} of the Wannier-Qi (WQ) basis.\cite{qi} Our approach has a number of advantages. First, it highlights the similarities with the standard FQH physics thus allowing for a direct comparison between the two.\cite{qi,moller,wannier,gaugefixing} Second, it provides a computationally efficient setting for a \textit{momentum-space} DMRG: For a lattice, whose linear dimension is $N$ unit cells, the logarithm of the computational cost on a cylinder scales as $N$, compared with $2N$ for DMRG on a torus, and $N^2$ for ED.\cite{gammatorus} Third, the orbital entanglement spectrum (OES) on cylinders probes the physics of a single edge, thus allowing for a cleaner and more straightforward identification of topological orders, compared to the torus setup which involves a nontrivial combination of two edges.\cite{Lauchli,Zhao} Fourth, it naturally allows for the inclusion of an external potential which is ubiquitous in possible cold-atom realizations and, as we find, helps to stabilize some of the fragile FCIs including the counterpart of the non-Abelian Moore-Read state.\cite{private} Finally, the presence of a physical boundary makes it possible to study the FCI edge states in detail.

Here we exploit all of these advantages by calculating the OES and the edge excitation spectrum for bosonic FCIs at filling fractions $\nu=1/2$ and $\nu=1$. We find that the OES in these systems have the same low-lying counting structure as the edge excitation spectrum, thus providing compelling evidence for the bulk-edge correspondence, similar to the one found in FQH states. In addition, we demonstrate that the Moore-Read FCI state at $\nu=1$, which in a standard setup requires three-body interactions on a torus, is likely to survive for more realistic two-body interactions in the presence of a parabolic potential, providing another possible way to realize non-Abelian phases which is different from the optical flux lattice setup of Refs.~\onlinecite{moessner,cooper}. We note that our calculations of the OES have been performed for systems which are much larger than the current limit of ED.

{\it Setup.}~We start with the lattice Hamiltonian on a finite torus whose periods are defined by two vectors $\textbf{v}_{1,2}$ with $N_{1,2}$ unit cells in the $\textbf{v}_{1,2}$ direction. The system consists of $N_b$ interacting bosons partially filling the lowest Bloch band of the ruby \cite{ruby} or the kagome \cite{chernins1} lattice [we adopt the same symbols for the hopping parameters as used in Ref.~\onlinecite{nonab3} and fix their values to $\{t_r,t_i,t_{1r},t_{1i},t_4\}=\{1,1.2,-1.2,2.4,-1.46\}$ for the ruby lattice, and $\{t_1,\lambda_1,t_2,\lambda_2\}=\{1,1,0,0\}$ for the kagome lattice]. We form a complete set of eigenstates in the lowest Chern band using gauge-fixed WQ orbitals localized in the~$\textbf{v}_1$ direction.\cite{qi,gaugefixing} The latter are counterparts of the lowest Landau level single-particle wave functions in the Landau gauge. This construction clarifies the connection between the lattice CIs and the QH systems. A generic translationally invariant two-body lattice interaction (projected to the lowest band in a standard way \cite{cherninsnum2}) has the following general form in the WQ basis
\begin{equation}\label{ham_FQH}
\hat{H}^{\textrm{tor}}_{\textrm{lat}}=\sum_{\{j_n\}=0}^{N_1N_2-1}\delta_{j_1+j_2,j_3+j_4}^{\textrm{mod}\  N_2}
V_{\{j_n\}}^{\textrm{lat,tor}}\hat{a}_{j_1}^\dagger \hat{a}_{j_2}^\dagger \hat{a}_{j_3} \hat{a}_{j_4},
\end{equation} where $\hat{a}_j^\dagger$ creates a boson in $j$'s orbital and there are $N_1N_2$ WQ orbitals on the torus. In order to extend this construction to the cylinder geometry with $N_s$ WQ orbitals ($N_s\ll N_1N_2$), we keep $N_2$ fixed while increasing $N_1$ until we reach the convergence of torus matrix elements $V_{\{j_n\}}^{\textrm{lat,tor}}$ with $j_n\in[0,N_s-1]$. This simple procedure generates a lattice with $N_s$ WQ orbitals on the finite-length cylinder. The lattice size in the $\textbf{v}_{2}$ direction is still $N_2$ (the same as that on the torus), while we have $N_1^{\textrm{cyl}}$ unit cells in the $\textbf{v}_{1}$ direction, where $N_1^{\textrm{cyl}}=N_s/N_2$. This system is described by the Hamiltonian $\hat{H}^{\textrm{cyl}}_{\textrm{lat}}$, which can be obtained from Eq.~(\ref{ham_FQH}) by substituting $N_1^{\textrm{cyl}}$ and the interaction matrix elements
$V_{\{j_n\}}^{\textrm{lat,cyl}}=\lim_{N_1\rightarrow \infty} V_{\{j_n\}}^{\textrm{lat,tor}}$.
The filling fraction $\nu$ is defined in terms of the number of particles, $N_b$, and the number of "flux quanta", $N_s=N_1^{\textrm{cyl}}N_2$, as $\nu=N_b/(N_s+\mathcal S)$, where $\mathcal S$ denotes system-size-independent integer ``shift,''  which is a topological quantum number characterizing FQH states in a finite geometry on a sphere or a cylinder, which  slightly increases the particle density.
Compared with the Hamiltonian of a FQH system on a finite cylinder the total momentum in our setup $K=\sum_{n=1}^{N_b} j_n$ is conserved only mod $N_2$.

Having constructed the Hamiltonian $\hat{H}^{\textrm{cyl}}_{\textrm{lat}}$, we use DMRG and ED to study the FCIs numerically. Our DMRG implementation is similar to the approach of Ref.~\onlinecite{kovrizhin}. The orbital entanglement spectrum of the ground state, which in the FQH case reflects the nature of edge excitations, can naturally be generated in finite-size DMRG sweeps. Using a standard recipe developed for the FQH systems \cite{orbitalcut} we partition WQ orbitals into two disjoint sets $A$ and $B$, consisting of $l_A$ consecutive orbitals with ``momentum'' $j$ running from $0$ to $l_A-1$ and the remaining $N_s-l_A$ orbitals with ``momentum'' $l_A$ to $N_s-1$. Generalized (mod $N_2$) total momentum conservation requires that each OES level is labeled by $N_A$ and $J_A\equiv K_A [\textrm{mod}\ N_2]$. Below we study ruby and kagome lattices of dimension $N_1^{\textrm{cyl}}\times N_2$ on finite cylinders whose QH counterpart is a cylinder with a circumference $L=l_BN_2\sqrt{2\pi/\sin(\pi/3)}$, where $l_B$ is the magnetic length.

\begin{figure}
\centerline{\includegraphics[width=\linewidth]{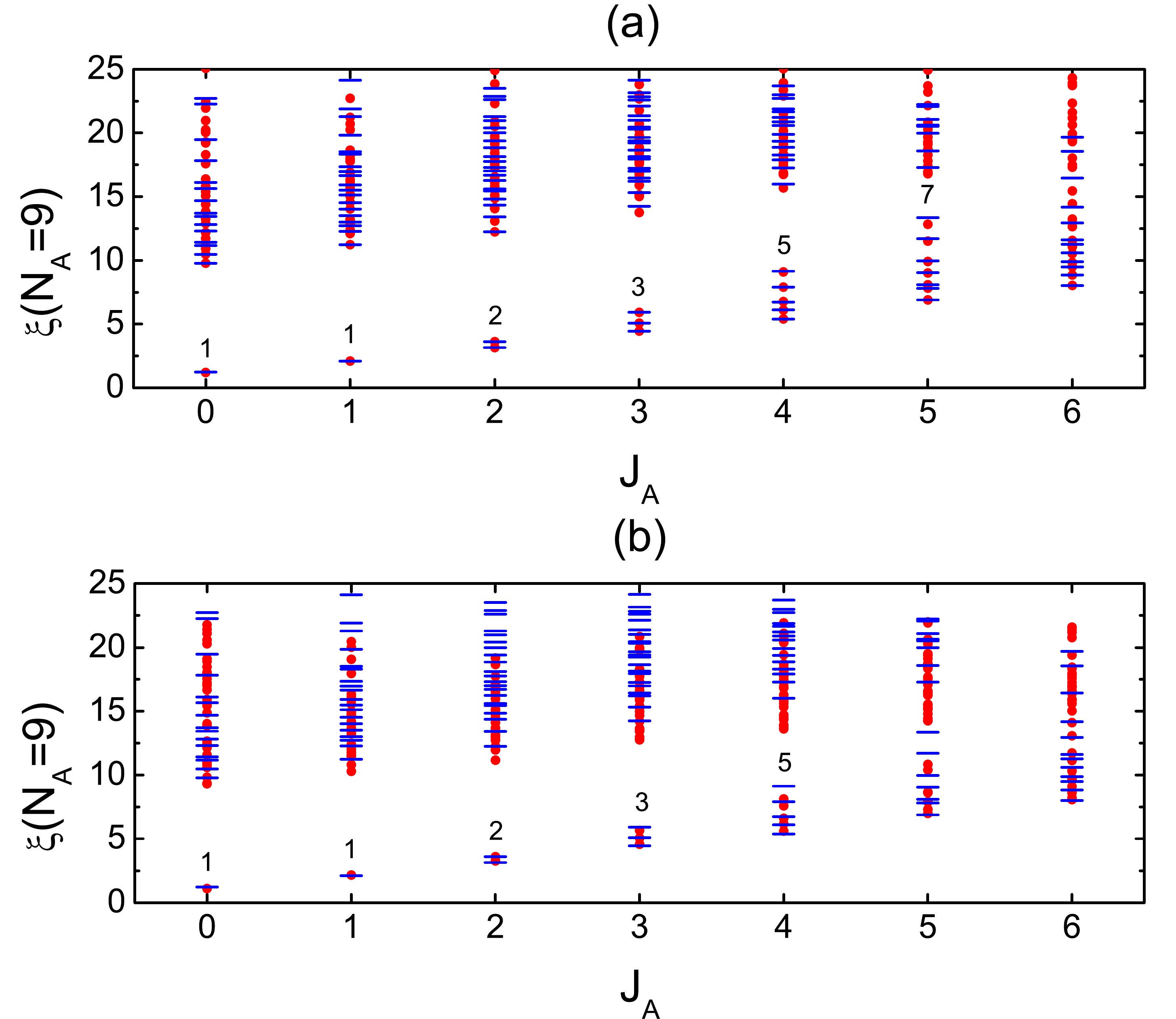}}
\caption{(Color online) The ground-state orbital entanglement spectrum for FCIs (red dots)
and the corresponding OES for the Laughlin state (blue dashed lines) from DMRG for $18$ bosons at $\nu=1/2$.
(a) The ground-state OES for $5\times7$ ruby lattice is obtained for DMRG cutoff $\varepsilon=10^{-10}$ after $13$ sweeps.
(b) The ground-state OES for $5\times7$ kagome lattice ($\varepsilon=10^{-9}$ after $10$ sweeps).
The OES of the Laughlin state is obtained with $\varepsilon=10^{-10}$ after $20$ sweeps.
\label{es_fci}}
\end{figure}

{\it Entanglement spectrum.}~Let us first focus on the case of the filling fraction $\nu=1/2$ in the presence of two-body on-site interactions $\sum_i \hat{n}_i(\hat{n}_i-1)$, where $\hat{n}_i$ is the number of particles on $i$'s lattice site.
Taking into account that the standard $\nu=1/2$ bosonic Laughlin state on a cylinder appears at $N_s=2 N_b-1$, we choose the same parameters for the FCI. Our ED calculation for $\hat{H}^{\textrm{cyl}}_{\textrm{lat}}$ on small systems shows a unique ground state with excited states separated by the many-particle gap. The overlap $|\langle\Psi_{\textrm{GS}}|\Psi_{\textrm{Lau}}\rangle|$ between the ground state and the Laughlin $\nu=1/2$ state reaches $0.9996$ and $0.9833$ for $8$ bosons on $3\times5$ ruby and kagome lattices correspondingly. These strong overlaps signal that the ground states of $\hat{H}^{\textrm{cyl}}_{\textrm{lat}}$ are in the same class as the Laughlin state.

To further identify the nature of topological orders in these systems, we use DMRG to calculate
the OES for a cut in the WQ basis. By analogy with the FQH case the OES is expected to reflect the properties of edge excitations. We choose $l_A=(N_s+1)/2$ and label each OES level by the particle number $N_A$ (displayed data is typically chosen so that $N_A$ corresponds to the pertinent root configuration) and the quasi-momentum $J_A$ of the $A$ set. The ground-state OES is presented in Fig.~\ref{es_fci}, where we have also included the results for the Laughlin state. Compared to the usual OES picture for FQH states, the one of the FCIs on a cylinder is folded, reflecting the generalized momentum conservation. This does not alter the counting of the low-lying spectrum, and one can clearly see the $\{1,1,2,3,5,7\}$ structure (Fig.~\ref{es_fci}). The spectrum of the $\nu=1/2$ FCI agrees well with the corresponding Laughlin state.

\begin{figure}
\centerline{\includegraphics[width=\linewidth]{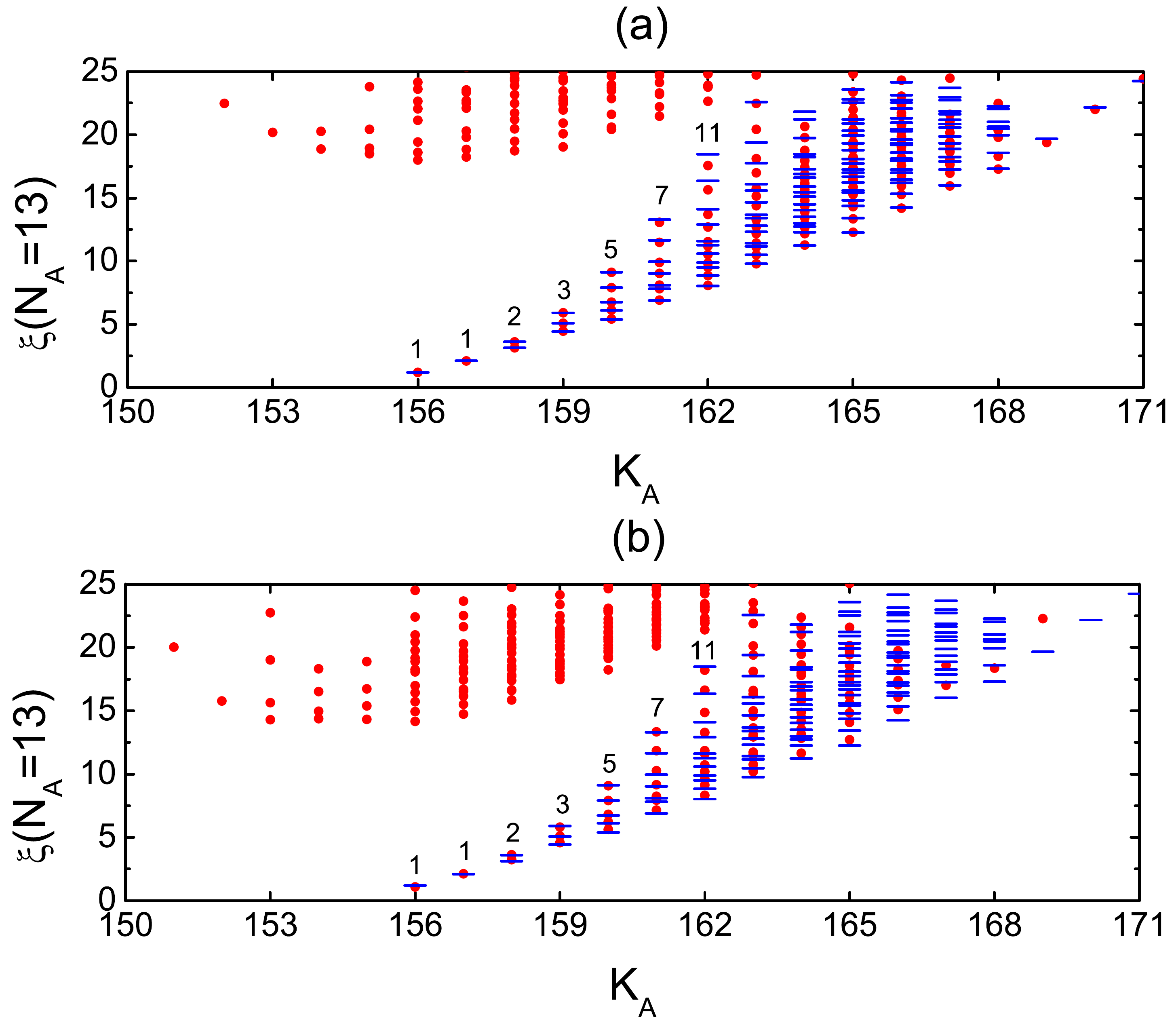}}
\caption{(Color online) The ground-state orbital entanglement spectrum for FCIs in the $K_\infty$ approximation (red dots) and the corresponding OES for the Laughlin state (blue lines) from DMRG for $25$ bosons at $\nu=1/2$. (a) The ground-state OES in the $K_\infty$ approximation for $7\times7$ ruby lattice ($\varepsilon=10^{-12}$ after $9$ sweeps).
(b) The ground-state OES in the $K_\infty$ approximation for $7\times7$ kagome lattice ($\varepsilon=10^{-11}$ after $11$ sweeps).
The OES of the Laughlin state is obtained with $\varepsilon=10^{-10}$ after $20$ sweeps.
\label{es_kfci}}
\end{figure}

Although the standard momentum $K$ conservation does not hold exactly in the finite cylinder geometry, the ED results show that the FCI ground states still have a large weight (reaching $99.94\%$ and $97.82\%$ for $8$ bosons on $3\times5$ ruby and kagome lattices) in the expected $K$ sector, i.e., $K=N_b(N_b-1)$ for a Laughlin state at $\nu=1/2$. We use this observation to justify the $K_\infty$ approximation; namely, in the following we omit the terms that break the standard $K$ conservation in the Hamiltonian $\hat{H}^{\textrm{cyl}}_{\textrm{lat}}$. Then, we can obtain the ground states with a fixed $K$ and label the OES by the standard $(N_A,K_A)$ pair (Fig.~\ref{es_kfci}). One can see that the low-lying part of the $K_\infty$ ground-state OES matches that of the $\nu=1/2$ Laughlin state. There is a clear entanglement gap $\Delta \xi$, defined as the difference between the lowest level and the first excited level in a fixed $(N_A,K_A)$ sector [$\Delta \xi\approx17$ for the Ruby lattice and $\Delta \xi\approx13.5$ for the Kagome lattice in the $(N_A=13,K_A=156)$ sector] which is larger than that for the $\nu=1/2$ Coulomb ground state in the FQH on a sphere ($\Delta \xi\approx10$ \cite{ronny}). These results, together with the folded ground-state OES, provide compelling evidence that the ground states of these systems at $\nu=1/2$ are in the same class as the FQH Laughlin state.

\begin{figure}
\centerline{\includegraphics[width=\linewidth]{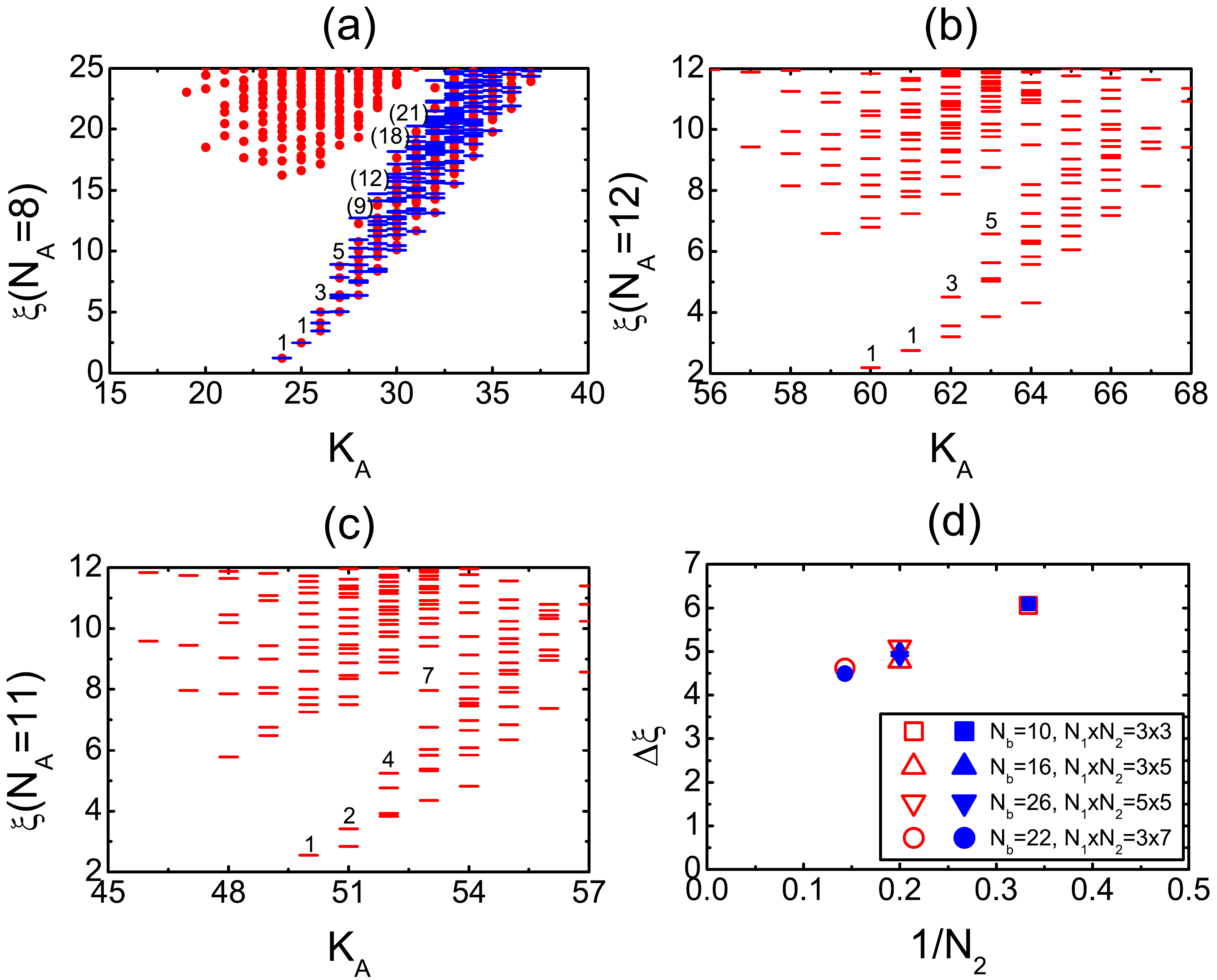}}
\caption{(Color online) (a) The ground-state orbital entanglement spectrum for FCIs in the $K_\infty$ approximation for $16$ bosons on $3\times5$ ruby lattice with three-body on-site interactions (red dots), compared with the OES of MR state (blue dashed lines) obtained from ED. Even the finite-size reduced conformal field theory (CFT) countings, indicated by numbers in parentheses, match identically in the FCI case. (b), (c) The ground-state OES in the $K_\infty$ approximation for $22$ bosons on $3\times7$ ruby lattice from DMRG for two-body on-site interactions with a confining potential $v_p=0.008$ and $N_{\textrm{trap}}=7$ ($\varepsilon=5\times10^{-8}$ after $18$ sweeps) in (b) $N_A=12$ and (c) $N_A=11$ sectors. (d) The finite-size scaling of the entanglement gap in $\Delta K_A=0$ sector (empty symbols) and $\Delta K_A=1$ sector (filled symbols) in the case of two-body interactions.
\label{es_pf}}
\end{figure}

Now let us consider the case of $\nu=1$ with $N_s=N_b-1$, where we will look for the FCI counterpart of the non-Abelian Moore-Read (MR) state. Compared with the Laughlin state, the MR-like phase in FCIs is more fragile, usually requiring three-body interactions on a torus.\cite{nonab1,nonab2,nonab3} Therefore, we first apply our cylinder setup to bosons with three-body on-site interactions $\sum_i \hat{n}_i(\hat{n}_i-1)(\hat{n}_i-2)$, where $\hat{n}_i$ is the number of particles on $i$'s lattice site. The ED result for 10 bosons on $3\times3$ Ruby lattice shows a unique ground state with a large overlap $|\langle\Psi_{\textrm{GS}}|\Psi_{\textrm{MR}}\rangle|=0.9997$ with the exact MR state and the weight $99.96\%$ in the expected $K$ sector, i.e., $K=N_b(N_b/2-1)$, which again justifies our $K_{\infty}$ approximation. We also obtain remarkable ED results for a larger system: The overlap between the ground state in the $K_\infty$ approximation and the exact MR state reaches $0.9998$ for $16$ bosons and the low-lying part of the ground-state OES matches that of the exact MR state with a very high precision and is accompanied by a large entanglement gap [$\Delta \xi\approx 15$ in the $(N_A=8,K_A=24)$ sector] [Fig. \ref{es_pf}(a)]. These results provide compelling evidence for the existence of FCIs with three-body interactions in the MR phase.

\begin{figure}
\centerline{\includegraphics[width=\linewidth]{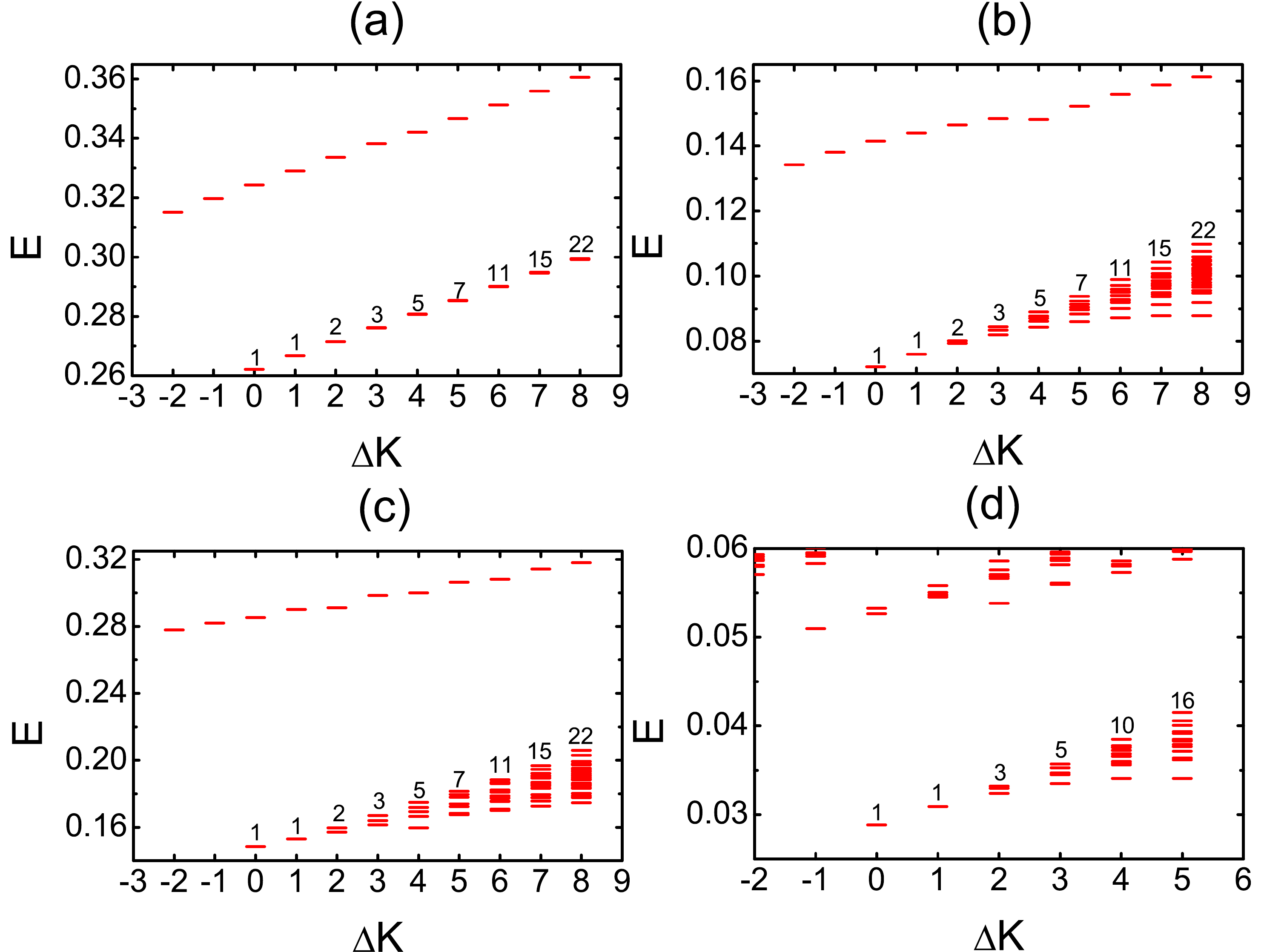}}
\caption{(Color online) Edge excitation spectrum from ED in the $K_\infty$ approximation at $\nu=1/2$ (two-body on-site interactions) and $\nu=1$ (three-body onsite interactions).
(a) $N_b=8$ on $5\times5$ ruby lattice with $v_l=0.01$.
(b) $N_b=8$ on $5\times5$ ruby lattice with $v_p'=0.0006$.
(c) $N_b=8$ on $5\times5$ kagome lattice with $v_p'=0.001$.
We observe the U(1) counting of low-energy excitations $\{1,1,2,3,5,7,...\}$ at $\Delta K=\{0,1,2,3,4,5,...\}$ in all three cases.
(d) $N_b=10$ on $5\times3$ ruby lattice with $v_p'=0.0002$.
We observe the U(1)$\times$Ising counting of low-energy excitations $\{1,1,3,5,10,16,...\}$ at $\Delta K=\{0,1,2,3,4,5,...\}$.
\label{edge_excitation}}
\end{figure}

Surprisingly, we find that the MR FCIs on cylinders can survive even in the case of more realistic two-body interactions and an additional parabolic confining potential $v_p\{\sum_{j=0}^{N_{\textrm{trap}}-1} [2\pi (j-N_{\textrm{trap}})/L]^2 \hat{a}_j^\dagger \hat{a}_j+\sum_{j=N_s-N_{\textrm{trap}}}^{N_s-1} [2\pi (j+N_{\textrm{trap}}-N_s+1)/L]^2 \hat{a}_j^\dagger \hat{a}_j\}$ acting on $N_{\textrm{trap}}$ WQ orbitals near the edges.\cite{private} The overlap between the ground state in the $K_\infty$ approximation and the exact MR state reaches $0.9138$ for $16$ bosons ($v_p=0.02, N_{\textrm{trap}}=5$), which is smaller than the three-body overlap, but is still highly nontrivial given that the Hilbert-space dimension in the relevant $K-$sector is large $\sim 3\times10^6$. The low-lying part of the OES with the $\{1,1,3,5\}$ and $\{1,2,4,7\}$ counting structures in two $N_A$ sectors, which is another signature of the MR phase, is also present in larger systems which we study using DMRG [Figs.~\ref{es_pf}(b), \ref{es_pf}(c)]. Finite-size scaling of the entanglement gap up to $26$ bosons in $\Delta K_A=0,1$ sectors ($N_A$ is chosen as the particle number in partition $A$ for the root configuration $2020...2$ of the MR state) reveals that it is governed by the circumference of the cylinder, $L\propto N_2$, whence $\Delta \xi(N_2)=\Delta\xi_\infty+\mathcal O(1/N_2)$. Figure~\ref{es_pf}(d) provides strong evidence that $\Delta\xi_\infty\neq 0$, i.e., that the gap between CFT levels and the generic levels indeed remains finite in the thermodynamic limit.

{\it Edge excitation spectrum.}~Open boundaries of finite cylinders provide
a natural setting for the studies of edge excitations which appear in the vicinity of real physical edges.
In our numerical approach we keep $N_2$ fixed and increase $N_1^{\textrm{cyl}}$ adding extra WQ orbitals, then open the edge on one side to allow occupation of these states while keeping the edge on the other side closed.
In order to observe a stable edge excitation spectrum, we consider various confining potentials that extend from the bulk
to the extra WQ orbitals. For weak potentials, a branch of low-energy excitations separated from higher levels appears in the spectrum for both filling fractions $\nu=1/2$ and $\nu=1$ as shown in Fig.~\ref{edge_excitation}. For a linear confinement, $v_l\sum_{j=0}^{N_s-1} (2\pi j/L) \hat{a}_j^\dagger \hat{a}_j$, the spectrum accurately matches the prediction of the Luttinger liquid theory:\cite{wenedge} The dispersion is linear and the edge states in each $\Delta K$ sector have nearly degenerate energies [Fig.~\ref{edge_excitation}(a)]. This degeneracy is lifted by a parabolic confinement, $v_p'\sum_{j=0}^{N_s-1} (2\pi j/L)^2 \hat{a}_j^\dagger \hat{a}_j$, which makes the excitation spectrum similar to the OES [Figs.~\ref{edge_excitation}(b)-\ref{edge_excitation}(d)].
The number of edge states in each $\Delta K$ sector does not depend on the form of the confinement and matches exactly with the conformal field theory prediction until the finite-size effects intervene at higher energies. Similar results have recently been obtained for a related problem of FQH states on a lattice in uniform magnetic field.\cite{kjall}

{\it Discussion.}~In summary, we studied bosonic fractional Chern insulators in the finite cylinder geometry using a combination of the exact diagonalization and the momentum-space DMRG. The ground-state OES at $\nu=1/2$ has a strong overlap with the OES of the corresponding Laughlin state. The ground-state OES at $\nu=1$ shows that the FCI analog of the FQH Moore-Read state is likely to survive even with two-body on-site interactions. The counting structure in the ground-state orbital entanglement, and the edge excitation spectrum, provides strong evidence for the bulk-edge correspondence in FCIs. Our setup is likely to bring new insights into intriguing and less understood FCI states which have no direct QH counterparts, most notably the states which can exist in flat bands with higher Chern numbers.\cite{ChernN,C2,hcnonab,cn}

Recently, similar results for the edge spectrum were obtained for the fractional Chern insulators in a disk geometry (using exact diagonalization in real space),\cite{yifeiedge} thus corroborating the universality of our findings.

{\it Acknowledgements.}~We acknowledge Andreas L\"auchli for valuable discussions.
E.~J.~B.~was supported by the Alexander von Humboldt foundation and by DFG's Emmy Noether program (BE 5233/1-1). Z.~L.~is
supported by China Postdoctoral Science Foundation Grant No. 2012M520149,
and acknowledges Ravindra Bhatt and Zi-Xiang~Hu for useful discussions. Z.~L. also thanks Hong-Gang Luo at Lanzhou Unversity
for the computational resources.
D.~K. is supported by EPSRC Grant No. EP/J017639/1 and EP/J009636/1, and acknowledges discussions with R.~Moessner and B.~Dou\c{c}ot.

\end{document}